\RequirePackage{amsmath}
\documentclass[runningheads]{llncs}

\usepackage[utf8]{inputenc}
\usepackage[T1]{fontenc}
\usepackage{lmodern}
\usepackage{microtype}
\usepackage[style=british]{csquotes}

\usepackage{amssymb}
\usepackage{mathtools}

\usepackage{tikz}
	\usetikzlibrary{matrix,arrows,shapes}
\usepackage{graphicx}
	
\usepackage{listings}
	\AtBeginDocument{\counterwithout{lstlisting}{section}}

\usepackage[%
	disable
]{todonotes}
\makeatletter
\RequirePackage[bookmarks,unicode,colorlinks=true]{hyperref}%
   \def\@citecolor{blue}%
   \def\@urlcolor{blue}%
   \def\@linkcolor{blue}%

\def\orcidID#1{\smash{\href{http://orcid.org/#1}{\protect\raisebox{-1.25pt}{\protect\includegraphics{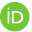}}}}}
\makeatother

\usepackage{xcolor}

\DeclareMathOperator{\Shift}{Shift}

\newcommand{\rb}{\rho_{\mathrm{b}}}
\newcommand{\rg}{\rho_{\mathrm{m}}}
\newcommand{\rgr}{\rho_{\mathrm{gr}}}
\newcommand{\Lb}{L_{\mathrm{b}}}
\newcommand{\LbOver}{\Lb}
\newcommand{\Kb}{K_{\mathrm{b}}}
\newcommand{\KbOver}{\Kb}
\newcommand{\Rb}{R_{\mathrm{b}}}
\newcommand{\RbOver}{\Rb}
\newcommand{\leb}{\mathit{le}_{\mathrm{b}}}
\newcommand{\rib}{\mathit{ri}_{\mathrm{b}}}
\newcommand{\mb}{m_{\mathrm{b}}}
\newcommand{\acb}{\mathit{ac}_{\mathrm{b}}}
\newcommand{\Lg}{L_{\mathrm{m}}}
\newcommand{\LgOver}{\Lg}
\newcommand{\Kg}{K_{\mathrm{m}}}
\newcommand{\KgOver}{\Kg}
\newcommand{\Rg}{R_{\mathrm{m}}}
\newcommand{\RgOver}{\Rg}
\newcommand{\leg}{\mathit{le}_{\mathrm{m}}}
\newcommand{\rig}{\mathit{ri}_{\mathrm{m}}}
\newcommand{\acg}{\mathit{ac}_{\mathrm{m}}}
\newcommand{\mg}{m_{\mathrm{m}}}

\newcommand{\iotaKg}{\iota_K^{\mathrm{m}}}

\newcommand{\rc}{\rho_{\mathrm{i}}}
\newcommand{\Lc}{L_{\mathrm{i}}}
\newcommand{\LcOver}{\Lc}
\newcommand{\Kc}{K_{\mathrm{i}}}
\newcommand{\KcOver}{\Kc}
\newcommand{\Rc}{R_{\mathrm{i}}}
\newcommand{\RcOver}{\Rc}
\newcommand{\lec}{\mathit{le}_{\mathrm{i}}}
\newcommand{\ric}{\mathit{ri}_{\mathrm{i}}}
\newcommand{\acc}{\mathit{ac}_{\mathrm{i}}}
\newcommand{\mc}{m_{\mathrm{i}}}
\newcommand{\nc}{n_{\mathrm{i}}}
\newcommand{\iotaLc}{\iota_L^{\mathrm{i}}}
\newcommand{\iotaKc}{\iota_K^{\mathrm{i}}}
\newcommand{\iotaRc}{\iota_R^{\mathrm{i}}}
\newcommand{\roo}{\rho_{\mathrm{e}}}
\newcommand{\rooComplete}{\roo = (\rb,\allowbreak \rg,\allowbreak \iota)}

\newcommand{\rcPrime}{\rc^{\prime}}
\newcommand{\mbPrime}{\mb^{\prime}}
\newcommand{\mcPrime}{\mc^{\prime}}

\newcommand{\type}[1]{\mathit{type}_{#1}}
\newcommand{\TG}{\mathit{TG}}
\newcommand{\src}[1]{\mathit{src}_{#1}}
\newcommand{\tar}[1]{\mathit{tar}_{#1}}

\newcommand{\Lover}{L}
\newcommand{\K}{K}
\newcommand{\R}{R}
\newcommand{\leit}{\mathit{le}}
\newcommand{\ri}{\mathit{ri}}
\newcommand{\Hover}{H}
\newcommand{\HPrimeOver}{H^{\prime}}
\newcommand{\D}{D}

\spnewtheorem{fact}[theorem]{Fact}{\bfseries}{\itshape}
\spnewtheorem*{assumption}{Assumption}{\bfseries}{\normalfont}

\providecommand{\longv}[1]{#1}
\providecommand{\shortv}[1]{}

\begin{document}

	\title{Finding the Right Way to Rome:\texorpdfstring{\\}{\ }
	Effect-oriented Graph Transformation}
	\longv{\subtitle{Extended Version}}
	\author{%
		Jens Kosiol\inst{1}\orcidID{0000-0003-4733-2777}\and
		Daniel Strüber\inst{2,3}\orcidID{0000-0002-5969-3521}\and
		Gabriele Taentzer\inst{1}\orcidID{0000-0002-3975-5238}\and
		Steffen Zschaler\inst{4}\orcidID{0000-0001-9062-6637}
	}

	\institute{%
		Philipps-Universität Marburg, Marburg, Germany\\
		\email{\{taentzer,kosiolje\}@mathematik.uni-marburg.de} 
		\and
		Chalmers $|$ University of Gothenburg, Gothenburg, Sweden\\ 
		\email{danstru@chalmers.se}
		\and
		Radboud University, Nijmegen, Netherlands\\
		\and
		King\rq{}s College London, London, UK\\
		\email{szschaler@acm.org}
	}
	
	\titlerunning{}
	\authorrunning{J. Kosiol, D. Strüber, et al.}

	\maketitle

	\begin{abstract}
    Many applications of graph transformation require rules that change a graph without introducing new consistency violations. 
    When designing such rules, it is natural to think about the desired outcome state, i.e., the desired \emph{effect}, rather than the specific steps required to achieve it; these steps may vary depending on the specific rule-application context. 
    Existing graph-transformation approaches either require a separate rule to be written for every possible application context or lack the ability to constrain the maximal change that a rule will create. 
    We introduce \emph{effect-oriented graph transformation}, shifting the semantics of a rule from specifying actions to representing the desired effect. 
    A single effect-oriented rule can encode a large number of \emph{induced} classic rules. 
    Which of the \emph{potential} actions is executed depends on the application context; ultimately, all ways lead to Rome. 
		If a graph element to be deleted (created) by a potential action is already absent (present), this action need not be performed because the desired outcome is already present. 
		We formally define effect-oriented graph transformation, show how matches can be computed without explicitly enumerating all induced classic rules, and report on a prototypical implementation of effect-oriented graph transformation in Henshin. 
		
		\keywords{Graph transformation \and Double-pushout approach \and Con\-sis\-ten\-cy-preserving transformations}
	\end{abstract}

	\section{Introduction}
	\label{sec:introduction}
Applications of graph transformation such as model synchronisation~\cite{FKST21,Fritsche22,Kosiol22} or search-based optimisation~\cite{Burdusel+21,HSBMZ22} require graph-transformation rules that combine a change to the graph with repair~\cite{SH19} operations to ensure transformations are consistency sustaining or even improving~\cite{Kosiol+21}.
For any given graph constraint, there are typically many different ways in which it can be violated, requiring slightly different specific changes to repair the violation.
As a result, it is often easier to think about the desired \emph{effect} of a repairing graph transformation rule rather than the specific transformations required. 
We would like to be able to reach a certain state of the graph---defined in terms of the presence or absence of particular graph elements (the \emph{effect})---even if, in different situations, a different set of specific changes is required to achieve this.

Existing approaches to graph transformation make it difficult to precisely capture the effect of a rule without explicitly specifying the specific set of changes required.
For example, the \emph{double-pushout approach (DPO)} to graph transformation~\cite{CMREHL97,EEPT06} has gained acceptance as an underlying formal semantics for \emph{graph} and \emph{model transformation rules} in practice as a simple and intuitive approach:
A transformation rule simply specifies which graph elements are to be deleted and created when it is applied; that is, a rule prescribes exactly all the \emph{actions} to be performed. 
For graph repair, this effectively forces one to specify every way in which a constraint can be violated and the specific changes to apply in this case, so that the right rule can be applied depending on context. 

On the other end of the spectrum, the \emph{double-pullback approach}~\cite{HEWC01} is much more flexible. 
Here, rules only specify minimal changes. 
However, there is no way of operationally constraining the maximal possible change. 

There currently exists no approach to graph transformation that allows the \emph{effect} of rules to be specified concisely and precisely without specifying every action that needs to be taken. 
In this paper, we introduce the notion of \emph{effect-oriented graph transformations.}
In this approach, graph-transformation rules encode a, potentially large, number of \emph{induced rules}.
This is achieved by differentiating basic actions that have to be performed by any transformation consistent with the rule and \emph{potential} actions that only have to be performed if they are required to achieve the intended rule \emph{effect.}
Depending on the application context, a different set of actions will be executed---all ways lead to Rome.
We provide an algorithm for selecting the right set of actions depending on context, without having to explicitly enumerate all possibilities---we efficiently find the right way to Rome. 

Thus, the paper makes the following contributions:
\begin{enumerate}
	\item We define the new notion of effect-oriented graph transformation rules and discuss different notions of consistent matches for these;
	\item We provide an algorithm for constructing a complete match and a transformation given a partial match for an effect-oriented transformation rule. The algorithm is efficient in the sense that it avoids computing and matching all induced rules explicitly; 
	\item We report on a prototypical implementation of effect-oriented transformations in Henshin; and 
  \item We compare our approach to existing approaches to graph transformation showing that it does indeed provide new expressivity.
\end{enumerate}

The rest of this paper is structured as follows. 
First, we introduce a running example (Sect.~\ref{sec:running-example}) and briefly recall basic preliminaries (Sect.~\ref{sec:preliminaries}). 
Section~\ref{sec:construction-effect-oriented-rules} introduces effect-oriented rules and transformations and several notions of constructing matches. 
Section~\ref{sec:locally-complete-matches} explains in more detail one algorithm for constructing matches for effect-oriented rules and reports on a prototype implementation in Henshin. 
In Section~\ref{sec:formal-comparison-loose-matching}, we discuss how existing applications can benefit from effect-oriented rules and transformations, and compare our new approach to graph transformation with other approaches that could be used to achieve these goals.
We conclude in Sect.~\ref{sec:conclusion}. 
\longv{In two appendices, we provide some additional results and explanations (Appendix~\ref{sec:additional-results}) and proofs of our formal statements (Appendix~\ref{sec:additional-proofs}).}
\shortv{We provide some additional results and explanations and proofs of our formal statements in an extended version of this paper~\cite{KSTZ23}.}

	\section{Running Example}
	\label{sec:running-example}
	We use the well-known banking example~\cite{Krause23} and adapt it slightly to illustrate our newly introduced concept of effect-oriented transformations.
Assume that the context of this example is specified in a meta-model formalised as a type graph (not shown) for the banking domain in which a \textsf{Bank} has \textsf{Clients}, \textsf{Accounts} and \textsf{Portfolios}. 
A \textsf{Client} may have \textsf{Accounts} which may be associated with a \textsf{Portfolio}. 

\begin{figure}[tbp]
	\centering
		\includegraphics[width=1.00\textwidth]{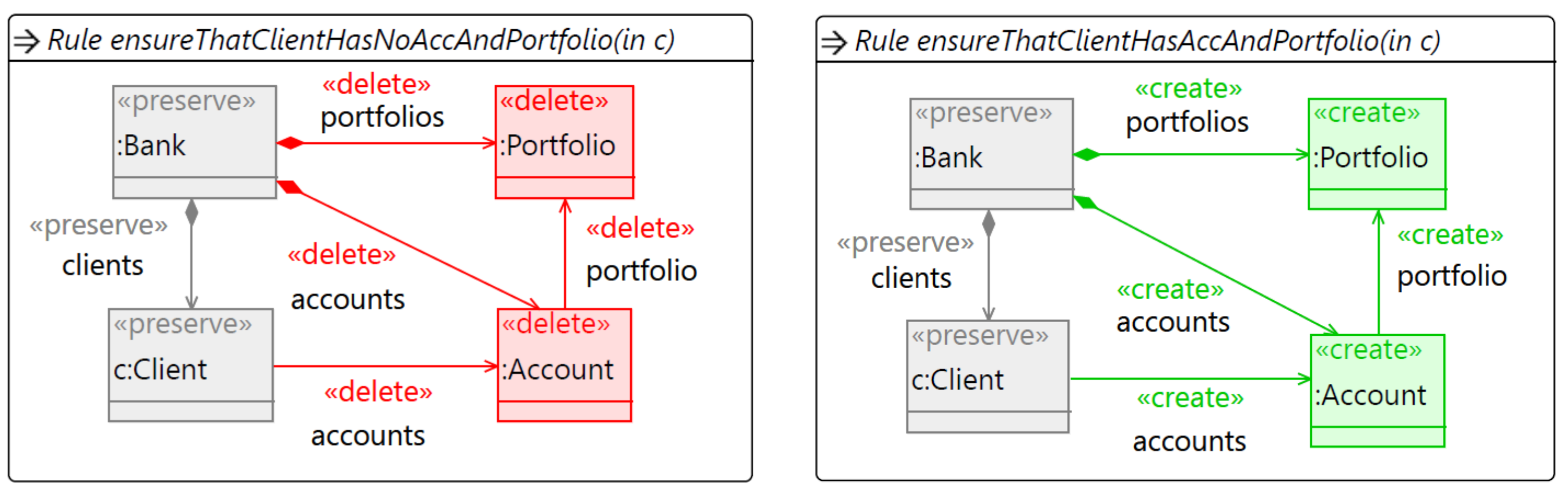}
	\caption{%
    Example rules, shown in the integrated visual syntax of Henshin~\cite{ABJKT10,struber2017henshin}.
    The LHS of a rule consists of all red and grey elements (additionally annotated with \textsf{delete} or \textsf{preserve}), the RHS of all  green (additionally annotated as \textsf{create}) and grey elements.}
	\label{fig:example-rules}
\end{figure}

Imagine a scenario where it is to be ensured that a \textsf{Client} has an \textsf{Account} with a \textsf{Portfolio}. 
To realise this condition in a rule-based manner so that no unnecessary elements are created,  at least three rules (and a programme to coordinate their application) are required:  
A rule that checks whether a \textsf{Client} already has an \textsf{Account} with a \textsf{Portfolio}, a rule that adds a \textsf{Portfolio} to an existing \textsf{Account} of the \textsf{Client}, and a rule that creates all the required structure; this last rule is shown as the rule \textsf{ensureThatClientHasAccAndPortfolio} in Fig.~\ref{fig:example-rules}. 

An analogous problem exists if the \textsf{Accounts} and \textsf{Portfolios} of a \textsf{Client} are to be removed. 
The rule \textsf{ensureThatClientHasNoAccAndPortfolio} in Fig.~\ref{fig:example-rules} is the rule that deletes the entire structure, and additional rules are needed to delete \textsf{Accounts} that are not associated with \textsf{Portfolios}. 
In general, the number of rules needed and the complexity of their coordination depend on the size of the structure to occur together and hence,  to be created (deleted). 

With our new notion of \emph{effect-oriented rules} and \emph{transformations}, we provide the possibility to use a single rule to specify \emph{all} desired behaviours by making the rule\rq{}s semantics dependent on the context in which it is applied. 
Specifically, if the rule \textsf{ensureThatClientHasAccAndPortfolio} is applied to a \textsf{Client} in effect-oriented semantics, this  allows for matching an existing \textsf{Account} and/or \textsf{Portfolio} (rather than creating them) and only creating the remainder. 
Therefore, we call the creation actions for \textsf{Account} and \textsf{Portfolio} \emph{potential actions} that are only executed if the corresponding elements do not yet exist. 
Similarly, applying the rule \textsf{ensureThatClientHasNoAccAndPortfolio} in effect-oriented semantics allows deleting nothing (if the matched \textsf{Client} has no \textsf{Account} and no \textsf{Portfolio}) or only an \textsf{Account} (that does not have a \textsf{Portfolio}). 
So here the deletion actions are potential actions that are only executed if the corresponding elements are present. 
We will allow for some of the actions of an effect-oriented rule to be mandatory. 
Note that in a Henshin rule, we would need to provide additional annotation to differentiate mandatory from potential actions.
We will define different strategies for this kind of matching that may be appropriate for different application scenarios.

	\section{Preliminaries}
	\label{sec:preliminaries}
	In this section, we briefly recall basic preliminaries. 
Throughout our paper, we work with \emph{typed graphs} and leave the treatment of attribution and type inheritance to future work. 
For brevity, we omit the definitions of \emph{nested graph conditions} and their \emph{shift} along morphisms~\cite{HP09,EGHLO14}. 
We also omit basic notions from category theory; in particular, we omit standard facts about adhesive categories~\cite{LS05,EEPT06} (of which typed graphs are an example). 
While these are needed in our proofs, the core ideas in this work can be understood without their knowledge. 

\begin{definition}[Graph. Graph morphism]
	A \emph{graph} $G = (V_G,E_G,\src{G},\tar{G})$ consists of a set of \emph{nodes} (or \emph{vertices}) $V_G$, a set of \emph{edges} $E_G$, and \emph{source} and \emph{target functions} $\src{G},\tar{G}\colon E_G \to V_G$ that assign a \emph{source} and a \emph{target node} to each edge. 
	
	A \emph{graph morphism} $f = (f_V,f_E)$ from a graph $G$ to a graph $H$ is a pair of functions $f_V\colon V_G \to V_H$ and $f_E\colon E_G \to E_H$ that both commute with the source and target functions, i.e., such that $\src{H} \circ f_E = f_V \circ \src{G}$ and $\tar{H} \circ f_E = f_V \circ \tar{G}$. 
	A graph morphism is \emph{injective}/\emph{surjective}/\emph{bijective} if both $f_V$ and $f_E$ are. 
	We denote injective morphisms via a hooked arrow, i.e., as $f\colon G \hookrightarrow H$. 
\end{definition}

Typing helps equip graphs with meaning; a \emph{type graph} provides the available types for elements and morphisms assign the elements of \emph{typed graphs} to those. 

\begin{definition}[Type graph. Typed graph]
	Given a fixed graph $\TG$ (the \emph{type graph}), a \emph{typed graph} $G = (G,\type{G})$ (over $\TG$) consists of a graph $G$ and a morphism $\type{G}\colon G \to \TG$. 
	A \emph{typed morphism} $f\colon G \to \Hover$ between typed graphs $G$ and $\Hover$ (typed over the same type graph $\TG$) is a graph morphism that satisfies $\type{H} \circ f = \type{G}$.
\end{definition}
Throughout this paper, we assume all graphs to be typed over a given type graph and all morphisms to be typed morphisms. 
However, for (notational) simplicity, we let this typing be implicit and just speak of graphs and morphisms. 
Moreover, all considered graphs are finite. 

\begin{definition}[Rules and transformations]\label{def:rules}
	A \emph{rule} $\rho = (p,\mathit{ac})$ consists of a \emph{plain rule} $p$ and an \emph{application condition} $\mathit{ac}$. 
	The plain rule is a span of 
	injective morphisms of typed graphs $p = (\Lover \xhookleftarrow{\leit} \K \xhookrightarrow{\ri} \R)$; its graphs are called \emph{left-hand side} (LHS), \emph{interface}, and \emph{right-hand side} (RHS), respectively. 
	The application condition $\mathit{ac}$ is a nested condition~\cite{HP09} over $\Lover$. 

	 Given a rule $\rho = (\Lover \xhookleftarrow{\leit} \K \xhookrightarrow{\ri} \R,\mathit{ac})$ and an injective morphism $m\colon \Lover \hookrightarrow G$, a \emph{(direct) transformation} $G \Longrightarrow_{\rho,m} \Hover$ from $G$ to $\Hover$ (in the Double-Pushout approach) is given by the diagram in Fig.~\ref{fig:def-rule-based-transformation} where both squares are pushouts and $m$ satisfies the application condition $\mathit{ac}$, denoted as $m \models \mathit{ac}$. 
	If such a transformation exists, the morphism $m$ is called a \emph{match} and rule $\rho$ is \emph{applicable} at match $m$; in this case, $n$ is called the \emph{comatch} of the transformation. 
	An injective morphism $m\colon \Lover \hookrightarrow G$ with $m \models \mathit{ac}$ from the LHS of a rule to some graph $G$ is called a \emph{pre-match}.

	\begin{figure}[t]
		\centering
		\begin{tikzpicture}
			\matrix (m) [	matrix of math nodes,
										row sep=1.25em,
										column sep=1.25em,
										minimum width=1.25em]
			{
				\Lover	&								&	\K	&								& \R \\
								&	\mathrm{(PO)}	&			&	\mathrm{(PO)}	&	\\
				G				& 							& \D	& 							& \Hover \\};
			\path[-stealth]
				(m-1-3) edge [left hook->] node [above] {\scriptsize $\leit$} (m-1-1)
								edge [right hook->] node [above] {\scriptsize $\ri$} (m-1-5)
								edge [right hook->] node [left] {\scriptsize $d$} (m-3-3)
				(m-1-1) edge [right hook->] node [below,sloped] {\scriptsize $m \vDash \mathit{ac}$} (m-3-1)
				(m-1-5) edge [right hook->] node [right] {\scriptsize $n$} (m-3-5)
				(m-3-3) edge [left hook->] node [below] {\scriptsize $g$} (m-3-1)
								edge [right hook->] node [below] {\scriptsize $h$} (m-3-5);
		\end{tikzpicture}
		\caption{A rule-based transformation in the Double-Pushout approach}
		\label{fig:def-rule-based-transformation}
	\end{figure}
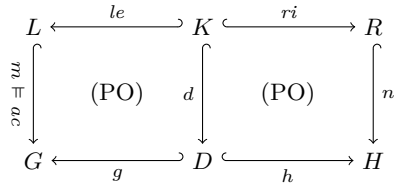
\end{definition}
For a rule to be applicable at a pre-match $m$, there must exist a pushout complement for $m \circ \leit$; in categories of graph-like structures, an elementary characterisation can be given in terms of the \emph{dangling condition}~\cite[Fact~3.11]{EEPT06}: 
A rule is applicable at a pre-match $m$ if and only if $m$ does not map a node to be deleted in $L$ to a node in $G$ with an incident edge that is not also to be deleted. 

Application conditions can be \enquote{shifted} along morphisms in a way that preserves their semantics~\cite[Lemma~3.11]{EGHLO14}. 
We presuppose this operation in our definition of \emph{subrules} without repeating it. 
Our notion of a subrule is a simplification of the concept of \emph{kernel} and \emph{multi-rules}~\cite{GHE14}.

\begin{definition}[Subrule]
	Given a rule $\rho = (\Lover \xhookleftarrow{\leit} \K \xhookrightarrow{\ri} \R,\mathit{ac})$, a \emph{subrule} of $\rho$ is a rule $\rho^{\prime} = (L^{\prime} \xhookleftarrow{\leit^{\prime}} K^{\prime} \xhookrightarrow{\ri^{\prime}} R^{\prime},\mathit{ac}^{\prime})$ together with a subrule embedding $\iota\colon \rho^{\prime} \hookrightarrow \rho$ where $\iota = (\iota_L,\iota_K,\iota_R)$ and $\iota_X\colon X^{\prime} \hookrightarrow X$ is an 
	injective morphism for $X \in \{L,K,R\}$ such that 
	both squares in Fig.~\ref{fig:definition-subrule} are pullbacks and $\mathit{ac} \equiv \Shift(\iota_L,\mathit{ac}^{\prime})$.
	
	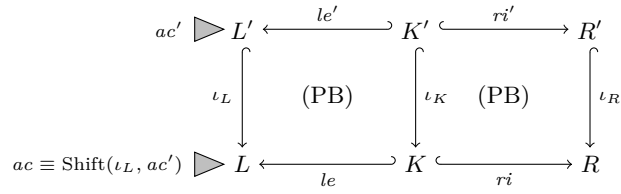
\begin{figure}
		\centering
		\begin{tikzpicture}
			\matrix (m) [	matrix of math nodes,
										row sep=1.25em,
										column sep=1.25em,
										minimum width=1.25em]
			{
				L^{\prime}	&								& K^{\prime}	&								& R^{\prime} \\
										&	\mathrm{(PB)}	&							&	\mathrm{(PB)}	&	\\
				\Lover			& 							& \K					& 							& \R \\};
			\node(d)[left of=m-1-1,node distance=4ex,isosceles triangle,draw=black,fill=lightgray,inner sep=2.5pt,label={[font=\scriptsize,label distance=1pt]180:$\mathit{ac}^{\prime}$}]{};
			\node(e)[left of=m-3-1,node distance=4ex,isosceles triangle,draw=black,fill=lightgray,inner sep=2.5pt,label={[font=\scriptsize,label distance=1pt]180:$\mathit{ac} \equiv \Shift(\iota_L,\mathit{ac}^{\prime})$}]{};
			\path[-stealth]
				(m-1-3) edge [left hook->] node [above] {\scriptsize $\leit^{\prime}$} (m-1-1)
								edge [right hook->] node [above] {\scriptsize $\ri^{\prime}$} (m-1-5)
								edge [right hook->] node [right] {\scriptsize $\iota_K$} (m-3-3)
				(m-1-1) edge [right hook->] node [left] {\scriptsize $\iota_L$} (m-3-1)
				(m-1-5) edge [right hook->] node [right] {\scriptsize $\iota_R$} (m-3-5)
				(m-3-3) edge [left hook->] node [below] {\scriptsize $\leit$} (m-3-1)
								edge [right hook->] node [below] {\scriptsize $\ri$} (m-3-5);
		\end{tikzpicture}
		\caption{Subrule $\rho^{\prime}$ of a rule $\rho$}
		\label{fig:definition-subrule}
	\end{figure}
\end{definition}

	\section{Effect-oriented Rules and Transformations}
	\label{sec:construction-effect-oriented-rules}
	The intuition behind effect-oriented semantics is that a rule prescribes the state that should prevail after its application, not the actions to be performed. 
In this section, we develop this approach. 
We introduce \emph{effect-oriented rules} as a compact way to represent a whole set of \emph{induced rules}. 
All induced rules share a common \emph{base rule} as subrule (prescribing actions to be definitively performed) but implement different choices of the \emph{potential actions} allowed by the effect-oriented rule. 
In a second step, we develop a semantics for effect-oriented rules; it depends on the larger context of effect-oriented transformations which of the induced rules is actually applied. 
Here, we implement the idea that potential deletions of an effect-oriented rule are to be performed if a suitable element exists but can otherwise be skipped.
In contrast, a potential creation is only to be performed if there is not yet a suitable element. 
This maximises the number of deletions to be made while minimising the number of creations.  
We propose two ways in which this \enquote{maximality} and \enquote{minimality} can be formally defined. 

\subsection{Effect-oriented Rules as Representations of Rule Sets}
In an \emph{effect-oriented rule}, a \emph{maximal rule} extends a \emph{base rule} by \emph{potential actions}. 
Here, and in all of the following, we assume that the left and right morphisms of rules and morphisms between rules (such as subrule embeddings) are actually inclusions. 
This does not lose generality (as the desired situation can always be achieved via renaming of elements) but significantly eases the presentation. 

\begin{definition}[Effect-oriented rule]\label{def:effect-oriented-rule}
	An \emph{effect-oriented rule} $\rooComplete$ is a rule $\rg = (\LgOver \xhookleftarrow{\leg} \KgOver \xhookrightarrow{\rig} \RgOver, \acg)$, called \emph{maximal rule}, together with a subrule $\rb = (\LbOver \xhookleftarrow{\leb} \KbOver \xhookrightarrow{\rib} \RbOver, \acb)$, called \emph{base rule}, and a subrule embedding $\iota\colon \rb \hookrightarrow \rg$ such that $\Kb = \Kg$ (and $\iota_K$ is an identity). 
	
	The \emph{potential deletions} of the maximal rule $\rg$ are the elements of 
		$(\Lg \setminus \Kg) \setminus \Lb = \Lg \setminus \Lb$; 
	analogously, its \emph{potential creations} are the elements of 
		$(\Rg \setminus \Kg) \setminus \Rb = \Rg \setminus \Rb$. 
	Here, and in the following, \enquote{$\setminus$} denotes the componentwise difference on the sets of nodes and edges. 
\end{definition}
While requiring $\Kb = \Kg$ restricts the expressiveness of effect-oriented rules, it suffices for our purposes and allows for simpler definitions of their matching. 
If, during matching, potential actions would compete with potential interface elements for elements to which they can be mapped, developing notions of maximality of matches becomes more involved. 

\begin{example}
	We consider the rules from Fig.~\ref{fig:example-rules} as the maximal rules of effect-oriented rules. 
	In each case, there are different possibilities as to which subrule of the rule to choose as the base rule. 
	Our convention that the interfaces of the base and maximal rules of an effect-oriented rule coincide specifies that in each case the base rule contains at least the interface that is to be preserved. 
	This minimal choice renders all deletions and creations potential. 
	
	Specifically, for the rule \textsf{ensureThatClientHasAccAndPortfolio}, one can assume that all elements to be created belong only to the maximal rule and represent potential creations. 
	A possible alternative is to consider as the base rule the rule that creates an \textsf{Account} (together with its incoming edges), making the creation of a \textsf{Portfolio} and its incoming edges potential. 
	Further combinations are possible.
\end{example}

An effect-oriented rule $\roo$ represents a set of \emph{induced rules}. 
The induced rules are constructed by extending the base rule of $\roo$ with potential deletions and creations from the maximal rule. 
However, we require every induced rule to have the same RHS as the maximal rule. 
Potential creations are omitted in an induced rule by also incorporating them into the interface. 
This ensures that the state that is represented by the RHS of the maximal rule holds after applying an induced rule, even if not all potential creations are performed. 
\begin{definition}[Induced rules]\label{def:induced-rules}
	Given an effect-oriented rule $\rooComplete$, every rule $\rc = (\LcOver \xhookleftarrow{\lec} \KcOver \xhookrightarrow{\ric} \RcOver, \acc)$ is one of its \emph{induced rules} if it is constructed in the following way (see (the upper part of) Fig.~\ref{fig:construction-and-matching-induced-rule}):
	\begin{enumerate}
		\item There is a factorisation $(1)$ $\iota_L = \iota_L^{\mathrm{m}\prime} \circ \iota_L^{\mathrm{i}\prime}$ of $\iota_L\colon \LbOver \hookrightarrow \LgOver$ into two inclusions $\iota_L^{\mathrm{m}\prime}$ and $\iota_L^{\mathrm{i}\prime}$. 
		
		\item There is a factorisation $(2)$ $\iota_R \circ \rib = \rig = \ric \circ \iotaKg$ of $\rig\colon \KgOver \hookrightarrow \RgOver$ into two inclusions $\ric$ and $\iotaKg$ such that the square $(2)$ is a pullback. 
		
		\item $(\LcOver, u,\lec)$ are computed as pushout of the pair of morphisms $(k_1,\iotaKg)$, where $k_1 \coloneqq \iota_L^{\mathrm{i}\prime} \circ \leb$; in that, we choose $\LcOver$ such that  $u$ and $\lec$ become inclusions (employing renaming if necessary). 
		
		\item The application condition $\acc$ is computed as $\Shift(\iotaLc,\acb)$, where $\iotaLc \coloneqq u \circ \iota_L^{\mathrm{i}\prime}$.
	\end{enumerate}
	
	\begin{figure}[t]%
		\centering
		\includegraphics[width=\textwidth]{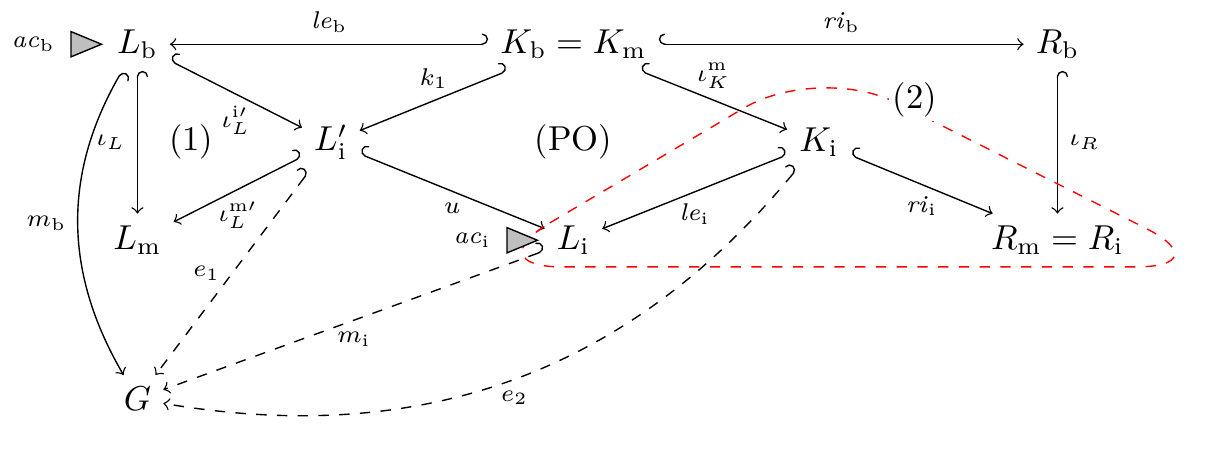}
		\caption{Construction of an induced rule $\rc$ (indicated via the red, dashed border) and of its match}
		\label{fig:construction-and-matching-induced-rule}
	\end{figure}
		
	The \emph{size} of an induced rule $\rc$ is defined as 
		$\lvert \rc \rvert \coloneqq \lvert L_{\mathrm{i}}^{\prime} \setminus \Lb \rvert + \lvert \Kc \setminus \Kb \rvert$.
\end{definition}

\begin{figure}[t]
	\centering
		\includegraphics[width=1.00\textwidth]{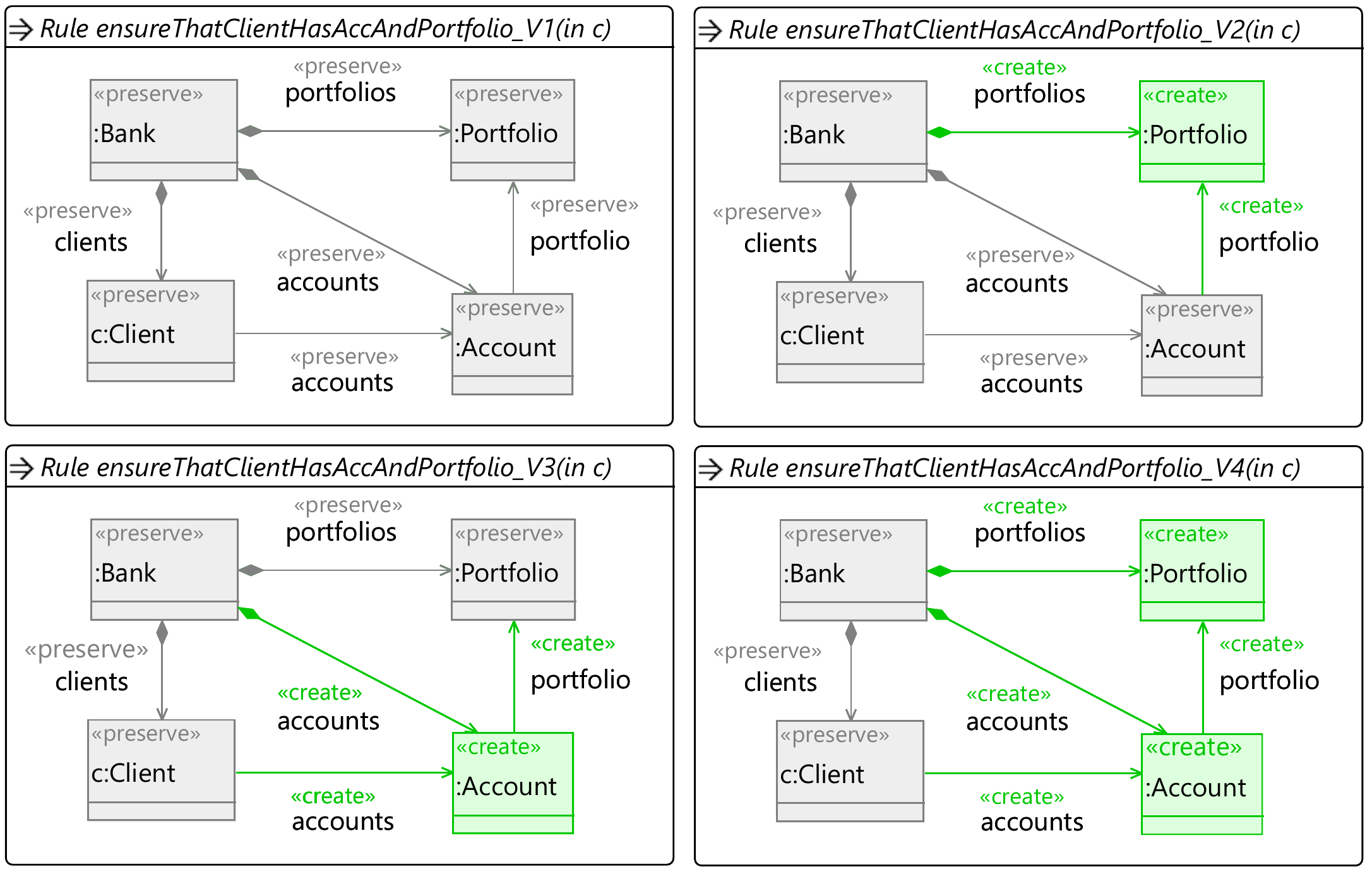}
	\caption{%
    Four induced rules arising from rule \textsf{ensureThatClientHasAccAndPortfolio}}
	\label{fig:example-induced-rules}
\end{figure}

\begin{example}\label{ex:induced-rules}
	Consider again rule \textsf{ensureThatClientHasAccAndPortfolio} as a maximal rule and its preserved elements as the base rule. 
	This effect-oriented rule has 26 induced rules, namely all rules that stereotype some of the \textsf{<<create>>}-elements of \textsf{ensureThatClientHasAccAndPortfolio} as \textsf{<<preserve>>}. 
	Figure~\ref{fig:example-induced-rules} shows a selection of those, namely induced rules that reduce undesired reuse of elements.
	These are the rules where, \textit{together with a node that is to be preserved (instead of being created), all adjacent edges that lead to preserved elements are also preserved.}
	\shortv{We call this the \textit{weak connectivity condition} and discuss and formalise it in~\cite{KSTZ23}.}
	\longv{We call this the \textit{weak connectivity condition} and discuss and formalise it in Appendix~\ref{sec:connectedness-conditions}.)}
	An example for an induced rule that is not depicted is the rule that creates a new \textsf{portfolio}-edge between existing \textsf{Accounts} and \textsf{Portfolios}. 
	
	Next, consider the effect-oriented rule where the maximal rule \textsf{ensureThatClientHasAccAndPortfolio} is combined with the base rule that already creates an \textsf{Ac\-count} with its two incoming edges. 
	Here,  the \textsf{Account} cannot become a context in any induced rule because its creation is already required by the base rule.  (This is ensured by the factorisation $(2)$ in Fig.~\ref{fig:construction-and-matching-induced-rule} being a pullback.) 
The induced rules of \textsf{ensureThatClientHasNoAccAndPortfolio} are obtained in a similar way. 
\end{example}

Our first result states that an induced rule actually contains the base rule of its effect-oriented rule as a subrule. 
In particular, this ensures that, if an induced rule is applied, all actions specified by the base rule are performed. 
\begin{proposition}[Base rule as subrule of induced rule]\label{prop:base-rule-as-subrule}
	If $\rc$ is an induced rule of the effect-oriented rule $\rooComplete$, then $\rb$ is a subrule of $\rc$ via the embedding $(\iotaLc,\iotaKc,\iotaRc)$, where $\iotaLc \coloneqq u \circ \iota_L^{\mathrm{i}\prime}$, $\iotaKc \coloneqq \iotaKg$ and $\iotaRc \coloneqq \iota_R$ (compare Fig.~\ref{fig:construction-and-matching-induced-rule}). 
\end{proposition} 

Next, we show that an effect-oriented rule indeed compactly represents a potentially large set of rules. 
The exact number of induced rules depends on how the edges of the maximal rule are connected. 
\begin{proposition}[Number of induced rules]\label{prop:number-induced-rules}
	For an effect-oriented rule $\rooComplete$, the number $n$ of its induced subrules (up to isomorphism) satisfies:
	\begin{equation*}
		2^{\lvert V_{\Lg} \setminus V_{\Lb} \rvert + \lvert V_{\Rg} \setminus V_{\Rb} \rvert} \leq n \leq 2^{\lvert V_{\Lg} \setminus V_{\Lb} \rvert + \lvert E_{\Lg} \setminus E_{\Lb} \rvert + \lvert V_{\Rg} \setminus V_{\Rb} \rvert + \lvert E_{\Rg} \setminus E_{\Rb} \rvert} \enspace .
	\end{equation*}
\end{proposition}

While our definition of induced rules is intentionally liberal, in many application cases it may be sensible to limit the kind of considered induced rules to avoid undesired reuse (e.g., connecting an \textsf{Account} to an already existing \textsf{Portfolio} of another \textsf{Client}). 
\shortv{In~\cite{KSTZ23}, we provide a definition that enables that.}
\longv{In Appendix~\ref{sec:connectedness-conditions}, we provide a definition that enables that.} 

\subsection{Matching Effect-oriented Rules}
In this section, we develop different ways to match effect-oriented rules. 
\emph{Effect-oriented transformations} get their semantics from \enquote{classical} Double-Pushout transformations using the induced rules of the applied effect-oriented rule.  
We will develop different ways in which an existing context determines which induced rule of an effect-oriented rule should be applied at which match. 

We assume a pre-match for the base rule of an effect-oriented rule to be given and try to to extend this pre-match to a match for an appropriate induced rule. 
The notion of \emph{compatibility} captures this extension relationship. 
\shortv{In~\cite{KSTZ23}, we present two technical lemmas that further characterise compatible matches.}
\longv{In Appendix~\ref{sec:compatible-matches}, we present two technical lemmas that further characterise compatible matches.}
\begin{definition}[Compatibility]\label{def:compatibility}
	Given an effect-oriented rule $\rooComplete$, a pre-match $\mb\colon \Lb \hookrightarrow G$ for its base rule and a match $\mc\colon \Lc \hookrightarrow G$ for one of its induced rules $\rc$ are \emph{compatible} if $\mc \circ \iotaLc = \mb$, where $\iotaLc = u \circ \iota_L^{\mathrm{i}\prime}\colon \Lb \hookrightarrow \Lc$ stems from the subrule embedding of $\rb$ into $\rc$ (compare Fig.~\ref{fig:construction-and-matching-induced-rule} and Proposition~\ref{prop:base-rule-as-subrule}). 
	
	An induced rule $\rc$ can be \emph{matched compatibly to $\mb$} if it has a match $\mc$ such that $\mb$ and $\mc$ are compatible. 
\end{definition} 

Given an effect-oriented rule and a pre-match for its base rule, there can be many different induced rules for which there is a compatible match. 
The following definition introduces different strategies for selecting such a rule and match, so that the corresponding applications form transformations that are complete in terms of deletion and creation actions to achieve the intended effect, which is why they are called effect-oriented transformations. 
Their common core is that in any effect-oriented transformation, a pre-match of a base rule is extended by potential creations and deletions from the maximal rule such that no further extension is possible. 
Intuitively, this ensures that all possible potential deletions but only necessary potential creations are performed (in a sense we make formally precise in Theorem~\ref{thm:characterising-eo-trafos}). 
A stricter notion is to maximise the number of reused elements. 

\begin{definition}[Local completeness. Maximality. Effect-oriented transformation]\label{def:oo-transformation}
	Given a pre-match $\mb\colon \LbOver \rightarrow G$ for the base rule $\rb$ of an effect-oriented rule $\rooComplete$, and a match $\mc$ for one of its induced rules $\rc$ that is compatible with $\mb$, $\rc$ and $\mc$ are \emph{locally complete} w.r.t. $\mb$ if (see Fig.~\ref{fig:construction-and-matching-induced-rule}):
	\begin{enumerate}
		\item \emph{Local completeness of additional deletions:} Any further factorisation $\iota_L = \iota_L^{\mathrm{m}\prime\prime} \circ \iota_L^{\mathrm{i}\prime\prime}$ into inclusions (with domain resp. co-domain $L_{\mathrm{i}}^{\prime\prime}$) such that there exists a non-bijective inclusion $j\colon L_{\mathrm{i}}^{\prime} \hookrightarrow L_{\mathrm{i}}^{\prime\prime}$ with $j \circ \iota_L^{\mathrm{i}\prime} = \iota_L^{\mathrm{i}\prime\prime}$ and $\iota_L^{\mathrm{m}\prime\prime} \circ j = \iota_L^{\mathrm{m}\prime}$ meets one of the following two criteria.
		\begin{itemize}
			\item \emph{Not matchable:} There is no injective morphism $e_1^{\prime}\colon L_{\mathrm{i}}^{\prime\prime} \hookrightarrow G$ with $e_1^{\prime} \circ \iota_L^{\mathrm{i}\prime\prime} = \mb$. 
			\item \emph{Not applicable:} Such an $e_1^{\prime}$ exists, but the morphism $\mc^{\prime}\colon \Lc^{\prime\prime} \to G$ which it induces together with the \emph{right extension match} $e_2$ of $\mc$ (where $\Lc^{\prime\prime}$ is the LHS of the induced rule that corresponds to this further factorisation) is not injective. 
		\end{itemize}	
		\item \emph{Local completeness of additional creations:} Any further factorisation $\rig = \ric^{\prime} \circ \iota_K^{\mathrm{m}\prime}$ into inclusions (with domain resp. co-domain $\Kc^{\prime}$) such that there is a non-bijective inclusion $j\colon \KcOver \hookrightarrow \Kc^{\prime}$ with $j \circ \iotaKg = \iota_K^{\mathrm{m}\prime}$ and $\ric^{\prime} \circ j = \ric$ meets one of the following two criteria. 
		\begin{itemize}
			\item \emph{Not matchable:} There is no injective morphism $e_2^{\prime}\colon \Kc^{\prime} \hookrightarrow G$ with $e_2^{\prime} \circ \iota_K^{\mathrm{m}\prime} = \mb \circ \leb$. 
			\item \emph{Not applicable:} Such an $e_2^{\prime}$ exists, but the morphism $\mc^{\prime}\colon \Lc^{\prime\prime} \to G$ which it induces together with the \emph{left extension match} $e_1$ of $\mc$ (where $\Lc^{\prime\prime}$ is the LHS of the induced rule that corresponds to this further factorisation) is not injective. 
		\end{itemize}
	\end{enumerate}

	An \emph{effect-oriented transformation} $t\colon G \Longrightarrow \Hover$ via $\roo$ is a double-pushout transformation $t\colon G \Longrightarrow_{\rc,\mc} \Hover$, where $\rc$ is an induced rule of $\roo$ and $\rc$ is locally complete w.r.t. $\mb\coloneqq \mc \circ \iotaLc$, the induced pre-match for $\rb$. 
	The \emph{semantics} of an effect-oriented rule is the collection of all of its effect-oriented transformations. 
	
	A transformation $t\colon G \Longrightarrow_{\rc,\mc} \Hover$ via an induced rule $\rc$ of a given effect-oriented rule $\roo$ is \emph{globally maximal} (w.r.t. $G$) if for any other transformation $t^{\prime}\colon G \Longrightarrow_{\rcPrime,\mcPrime} \HPrimeOver$ via an induced rule $\rcPrime$ of $\roo$,  it holds that $\lvert \rc \rvert \geq \lvert \rcPrime \rvert$. 
	Such a transformation $t$ is \emph{locally maximal} if for any other transformation $t^{\prime}\colon G \Longrightarrow_{\rcPrime,\mcPrime} \HPrimeOver$ via an induced rule $\rcPrime$ of $\roo$ where the induced pre-matches $\mb$ and $\mbPrime$ for the base rule coincide, it holds that $\lvert \rc \rvert \geq \lvert \rcPrime \rvert$. 
	In all of these situations, we also call the match $\mc$ and the rule $\rc$ \emph{locally complete} or \emph{locally}/\emph{globally maximal}.
\end{definition}

\begin{example}
	To illustrate the different kinds of matching for effect-oriented rules, we again consider \textsf{ensureThatClientHasAccAndPortfolio} as a maximal rule whose \textsf{<<preserve>>}-elements form the base rule and apply it according to different semantics to the example instance depicted in Fig.~\ref{fig:example-instance-minimal}. 
	First, we consider the base match that maps the \textsf{Client}-node of the rule to \textsf{Client c1} in the instance. 
	Extending this base match in a \emph{locally complete} fashion requires one to reuse the existing \textsf{Portfolio} and one of the existing \textsf{Accounts}. 
	Choosing \textsf{Account a2} leads to induced rule \textsf{ensureThatClientHasAccAndPortfolio\_V1} (Fig.~\ref{fig:example-induced-rules}) because there already exists an edge to \textsf{Portfolio p}. 
	In contrast, choosing \textsf{Account a1} leads to a transformation that creates a \textsf{portfolio}-edge from \textsf{a1} to \textsf{p} (where the underlying induced rule is not depicted in Fig.~\ref{fig:example-induced-rules}). 
	Both transformations are locally complete; in particular, locally complete matching is not deterministic. 
	If, for semantic reasons, one wants to avoid transformations like the second one and only allows the induced rules that are depicted in Fig.~\ref{fig:example-induced-rules}, applying \textsf{ensureThatClientHasAccAndPortfolio\_V2} at \textsf{Account a1} becomes locally complete. 
	
	The unique match for \textsf{ensureThatClientHasAccAndPortfolio\_V1} is the only locally maximal match compatible with the chosen base match in our example and is also globally maximal.
	To see a locally maximal match that is not globally maximal, we consider the base match that maps to \textsf{Client c2} instead of \textsf{c1}. 
	Here, the locally maximal match reuses \textsf{a2, p} and the \textsf{portfolio}-edge between them and creates the missing edges from \textsf{c2} to \textsf{a2} and \textsf{p}. 
	Choosing \textsf{Account a1} instead of \textsf{a2} does not provide a locally maximal match as one cannot reuse a \textsf{portfolio}-edge then (reducing the size of the induced rule by 1). 
	The globally maximal match remains unchanged as, by definition, it does not depend on a given base match. 
	It is that evident that in a larger example, every \textsf{Client} with an \textsf{Account} with \textsf{Portfolio} constitutes a globally maximal match. 
	Thus, also globally maximal matching is non-deterministic. 
	In fact, one can even construct examples where globally maximal matches for different induced rules (of equal size) exist.

	\begin{figure}[t]%
		\centering
		\includegraphics[width=\textwidth]{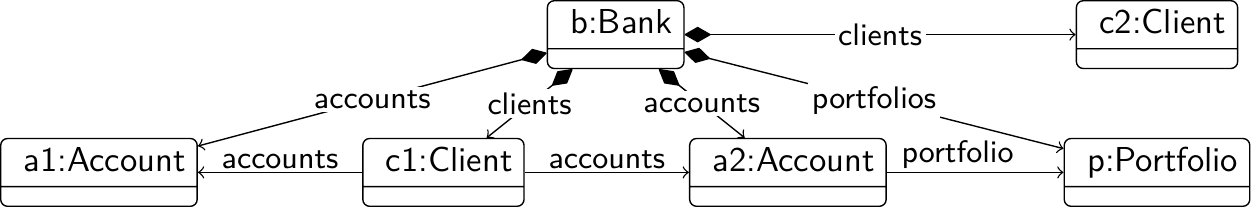}
		\caption{Tiny example instance for the banking domain}
		\label{fig:example-instance-minimal}
	\end{figure}
	
	Unlike potential creations, potential deletions may require backtracking to find a match. 
	To see this, consider the rule \textsf{ensureThatClientHasNoAccAndPortfolio} as a maximal rule whose \textsf{<<preserve>>}-elements form the base rule.
	Assuming that an \textsf{Account} can be connected to multiple \textsf{Clients}, a locally complete pre-match for an induced rule is not automatically a match for it.
	One has to look for an \textsf{Account} that is only connected to the matched \textsf{Client}.
\end{example}

The above example shows that none of the defined notions of transformation is deterministic. 
The situation is similar to Double-Pushout transformations in general, where the selection of the match is usually non-deterministic; however, in our case, there are different possible outcomes for the same base match.
For the applications we are aiming at, such as rule-based search, graph repair or model synchronisation (see Sect.~\ref{sec:effect-oriented-trafos-practice}), this is not a problem.
In these, it is often sufficient to know that, for instance, the selected \textsf{Client} has an \textsf{Account} and \textsf{Portfolio} after  applying the rule \textsf{ensureThatClientHasAccAndPortfolio} but not necessarily important which ones. 
 
It is easy to see that every globally maximal transformation via an induced rule is also locally maximal, and that every locally maximal transformation is locally complete. 
The definition of an effect-oriented transformation thus captures the weakest case and also covers locally and globally maximal transformations.

\begin{proposition}[Relations between different kinds of effect-oriented transformations]\label{prop:relations-concepts-transformation}
	Given an effect-oriented rule $\rooComplete$ and a graph $G$, every transformation $t\colon G \Longrightarrow_{\rc,\mc} \Hover$ via an induced rule $\rc$ of $\roo$ that is globally maximal is also locally maximal. 
	Every locally maximal transformation is also locally complete for its induced pre-match $\mb$ for the base rule $\rb$. 
\end{proposition}

Intuitively, the maximal rule of an effect-oriented rule specifies a selection of potential actions. 
The induced rules result from the different possible combinations of potential actions. 
The next theorem clarifies which effects can ultimately occur \emph{after} an effect-oriented transformation: 
If an element remains for which a matching potential deletion was specified by the effect-oriented rule, one of two alternatives took place: 
Either, the potential deletion was performed but on a different element (\emph{alternative action})---if there is more than one way to match an element potentially to be deleted. 
Or, that element was matched to by a potential creation (\emph{alternative creation}). 
The latter can happen when a rule specifies potential creations and deletions for elements of the same type (at comparable positions). 
Similarly, if $x$ denotes a performed potential creation but there had been an element $y$ to which $x$ could have been matched, $y$ was used by another potential creation or deletion (\emph{alternative action}). 
In particular, Theorem~\ref{thm:characterising-eo-trafos} also shows that effect-oriented rules can specify alternative actions. 
Their application can be non-deterministic, where one of several possible actions is chosen at random. 

\begin{theorem}[Characterising effect-oriented transformations]\label{thm:characterising-eo-trafos}
	Let $\rooComplete$ be an effect-oriented rule and $t\colon G \Longrightarrow_{\rc,\mc} H$ an effect-oriented transformation via one of its induced rules $\rc$ (compare Fig.~\ref{fig:construction-and-matching-induced-rule} for the following). 
	
	Let $x \in \Lg \setminus \Lb$ be an element that represents a potential deletion of $\roo$ and let $\Kb^{+}$ be the extension of $\Kb$ with $x$ (if defined as graph) and $\iota^+\colon \Kb \hookrightarrow \Kb^+$ the corresponding inclusion. 
	If there exists an injective morphism $m^+\colon \Kb^{+} \hookrightarrow H$ with $m^+ \circ \iota^+ = \nc \circ \iota_R \circ \rib$, where $\nc$ is the comatch of $t$, then either
	\begin{enumerate}
		\item \emph{(Alternative action):} the element $x$ belongs to $\Lc$; in particular, an element of the same type as $x$ (and in comparable position) was deleted from $G$ by $t$; or
		\item \emph{(Alternative creation):} the element $m^+(x)$ of $H$ has a pre-image from $\Rc$ under $\nc$, i.e., it was first created by $t$ or matched by a potential creation.
	\end{enumerate}
	
	Similarly, let $x \in \Rc \setminus (\Kc \cup \Rb)$ represent one of the potential creations of $\roo$ that have been performed by $t$. 
	Let $\Kb^+$ be the extension of $\Kb$ with $x$ (if defined as graph) and $\iota^+\colon \Kb \hookrightarrow \Kb^+$ the corresponding inclusion.  
	Then either
	\begin{enumerate}
		\item \emph{(Alternative action):} for every injective morphism $\mb^+\colon \Kb^+ \hookrightarrow G$ with $\mb^+\circ \iota^+ = \mc \circ \lec \circ \iotaKg$, the element $\mb^+(x) \in G$ has a pre-image from $\Lc$ under $\mc$ (i.e., it is already mapped to by another potential action); or 
		\item \emph{(Non-existence of match):} no  injective morphism $\mb^+\colon \Kb^+ \hookrightarrow G$ with $\mb^+\circ \iota^+ = \mc \circ \lec \circ \iotaKg$ exists.
	\end{enumerate}
\end{theorem}

	\section{Locally Complete Matches---Algorithm and Implementation}
	\label{sec:locally-complete-matches}
		In this section, we present an algorithm for the computation of a locally complete match (and a corresponding induced rule) from an effect-oriented rule and a pre-match for its base rule. 
Starting with such a pre-match is well-suited for practical applications such as model repair and rule-based search, where a match of the base rule is often already fixed and needs to be complemented by (some of) the actions of the maximal rule. 
Note that this pre-match is common to all matches of all induced rules. 
Searching for the pre-match once and extending it contextually is generally much more efficient than searching for the matches of the induced rules from scratch. 
Moreover, we simultaneously compute a locally complete match and its corresponding induced rule and thus avoid first computing all induced rules (of which there can be many, cf. Proposition~\ref{prop:number-induced-rules}) and trying to match them in order of their size. 
The correctness of our algorithm is shown in Theorem~\ref{thm:correctness-algorithm} below.
Note that we have focussed on the correctness of the algorithm and, apart from the basic efficiency consideration above, further optimisations for efficiency (incorporating ideas from~\cite{BP12}) are reserved for future work. 
\longv{We provide a short comment on our use of backtracking in Appendix~\ref{sec:representation-induced-rule}.} 
\shortv{We provide a short comment on our use of backtracking in~\cite{KSTZ23}.} 
In Sec.~\ref{sec:implementation} we report on a prototype implementation of effect-oriented transformations in Henshin using this algorithm.

\subsection{An Algorithm for Computing Locally Complete Matches}
\label{sec:matching-effect-oriented-rules}

We consider the problem of finding a locally complete match from a given pre-match. 
So-called \emph{rooted rules}, i.e., rules where a partial match is fixed (or at least can be determined in constant time), have been an important part of the development of rule-based algorithms for graphs that run in linear time~\cite{BP12,CCP22}. 
Note that we are not looking for an induced rule and match that lead to a maximal transformation but only to a non-extensible, i.e., locally complete one. 
This has the effect that the dangling-edge condition remains the only possible source of backtracking in the matching process. 

In Algorithm~\ref{alg:matching-induced-rules}, we outline a function that extends a pre-match for a base rule of an effect-oriented rule to a compatible, locally complete match for a corresponding induced rule. 
The input to our algorithm is an effect-oriented rule $\rooComplete$ (the parameter \emph{rule}), a graph $G$ (the parameter \emph{graph}) and a pre-match $\mb\colon \Lb \hookrightarrow G$ for $\rb$ (the parameter \emph{currentMappings}).
It returns a match $\mc$ for an induced rule $\rc$ of $\roo$ such that $\mc$ and $\mb$ are compatible and $\mc$ is locally complete; it returns \textsf{null} if and only if no such compatible, locally complete match exists. 
We outline the matching of nodes and consequently consider \emph{currentMappings} to be a list of node mappings; from this, one can infer the matching of edges. 
The computed match also represents the corresponding induced rule. 
\shortv{We provide the details for these conventions in~\cite{KSTZ23}.}
\longv{We provide the details for these conventions in Appendix~\ref{sec:representation-induced-rule}.}

The search for a match starts with initialising \emph{unboundNodes} with the potential actions of the given effect-oriented rule (line~5). 
Then the function \emph{findExtension} recursively tries to match those, extending the pre-match (line~6). 
Function \emph{findExtension} works as follows. 
For the unbound node $n$ at the currently considered \emph{position}, all available candidates, i.e., all nodes $x$ in the graph $G$ to which $n$ can be mapped, are collected with the function \emph{findExtensionCandidates} (line~10). 
A candidate $x$ must satisfy the following properties: (i) no other node may already map to $x$, i.e., $x$ does not yet occur in \emph{currentMappings} (injectivity condition) and (ii) the types of $n$ and $x$ must coincide (consistency condition). 
If candidates exist, for each candidate $x$, the algorithm tries to map $n$ to $x$ until a solution is found. 
To do this, the set of current mappings is extended by the pair $(n,x)$ (line~13). 
If $n$ was the last node to be matched (line~14) and the morphism defined by \emph{currentMappings} satisfies the dangling-edge condition (line~15), the result is returned as solution (line~16). 
If the dangling-edge condition is violated, the selected candidate is removed and the next one is tried (line~17). 
If $n$ was not the last node to be mapped (line~18), the function \textsf{findExtension} is called for the extended list of current mappings and the next unmatched node from \emph{unboundNodes} (line~19). 
If this leads to a valid solution, this solution is returned (lines~20--21).
Otherwise, the pair $(n,x)$ is removed from \emph{currentMappings} (line~23), and the next candidate is tried. 
If candidates exist but none of them lead to a valid solution, \textsf{null} is returned (line~33). 
If no candidate exists (line~25), either the current mapping or \textsf{null} is returned as the solution (if $n$ was the last node to assign; lines~26--29) or \emph{findExtension} is called for the next \emph{position} (line~31), i.e., the currently considered node $n$ is omitted from the mapping (and, hence, from the induced rule).

\begin{lstlisting}[caption={Computation of a locally complete match},label={alg:matching-induced-rules}]
input: effect-oriented rule $(\rb,\rg,\iota)$, graph $G$, and a pre-match $\mb$
output: locally complete match $\mc$ compatible with $\mb$

function findLocallyCompleteMatch(rule, graph, currentMappings)
	unboundNodes = $V_{\Lg} \setminus V_{\Lb} \sqcup V_{\Rg} \setminus V_{\Rb}$;
	return findExtension(currentMappings, graph, unboundNodes, 0);

function findExtension(currentMappings, graph, unboundNodes, position)
	n = unboundNodes.get(position);
	candidates = findExtensionCandidates(currentMappings, graph, n);
	if (!candidates.isEmpty())
		for each x in candidates
			currentMappings.put(n,x);
			if (position == unboundNodes.size() - 1) //last node to be matched
				if (danglingEdgeCheck(graph, currentMappings)) 
					return currentMappings;
				else currentMappings.remove(n,x);
			else //map next unbound node
				nextSolution = findExtension(currentMappings, graph, unboundNodes, position + 1);
				if (nextSolution != null)
					return nextSolution;
				else //try next candidate
				    currentMappings.remove(n,x);
		end for //no suitable candidate found
	else //there is no candidate for the current node
		if (position == unboundNodes.size() - 1) //last node to be matched
			if (danglingEdgeCheck(graph, currentMappings))
				return currentMappings;
			else return null;
		else //map next unbound node
			return findExtension(currentMappings, graph, unboundNodes, position+1);
	//no candidate led to a valid mapping
	return null;
\end{lstlisting}

\begin{theorem}[Correctness of Algorithm~\ref{alg:matching-induced-rules}]\label{thm:correctness-algorithm}
	Given an effect-oriented rule $\rooComplete$, a graph $G$, and a pre-match $\mb\colon \Lb \hookrightarrow G$, Algorithm~\ref{alg:matching-induced-rules} terminates and computes an induced rule $\rc$ with a match $\mc$ such that $\mc$ and $\mb$ are compatible and $\rc$ is locally complete w.r.t. $\mb$. 
	In particular, Algorithm~\ref{alg:matching-induced-rules} returns \textsf{null} if and only if no induced rule of $\roo$ can be matched compatibly with $\mb$ and returns $\rb$ as induced rule with match $\mb$ if and only if $\mb$ is locally complete and a match. 
\end{theorem}

		\subsection{Implementation}
		\label{sec:implementation}
		In this section, we present a prototypical implementation of effect-oriented transformations in Henshin~\cite{ABJKT10,struber2017henshin}, a  model transformation language based on graph transformation.
From the user perspective, the implementation includes two major classes for applying a given effect-oriented rule $\rooComplete$ to a host graph $G$:
the class \texttt{LocallyCompleteMatchFinder} for finding a locally complete match,  
and the class \texttt{EffectOrientedRuleApplication} to apply a rule $\rooComplete$ at such a match.
We assume that $\rooComplete$ is provided as simple Henshin rule representing the maximal rule $\rg$, where the base rule $\rb$ is implicitly represented by the preserved part of the rule.
The host graph $G$ is provided, as usual in Henshin, in the form of a model instance for a given meta-model (representing the type graph).

The implementation follows the algorithm presented in Section \ref{sec:matching-effect-oriented-rules}. 
Our main design goal was to reuse the existing interpreter core of Henshin, with its functionalities for matching and rule applications, as much as possible.
In particular, in \texttt{LocallyCompleteMatchFinder}, we derive the base rule $\rb$  by creating a copy of $\rg$ with creations and deletions  removed, and feeding it into the interpreter core to obtain a pre-match $\mb$ on $G$.
For cases where a pre-match $\mb$ can be found, we provide an implementation of Algorithm~\ref{alg:matching-induced-rules} that produces a partial match $\hat{\mg}$ incorporating the mappings of $\mb$ and additional mappings for elements to be deleted and elements not to be created. 
In order to treat elements of different actions consistently, we perform these steps on an intermediate rule $\rgr$, called the \textit{grayed} rule, in which creations are converted to preserve actions.
In \texttt{EffectOrientedRuleApplication}, we first derive the induced rule $\rc$ from  $\hat{\mg}$, such that $\hat{\mg}$ is a complete match for $\rc$. 
For the actual rule application, we feed $\rc$ together with $\hat{\mg}$ into the Henshin interpreter core using a classical rule application. 

We have tested the implementation using our running example. 
For this purpose, we specified all rules and an example graph. 
Our implementation behaved completely as expected. 
The source code of the implementation and the example are available online at \texttt{https://github.com/dstrueber/effect-oriented-gt}.

	\section{Related Work}
	\label{sec:formal-comparison-loose-matching}
	In this section, we describe how existing practical applications could benefit from the use of effect-oriented transformations (Sect.~\ref{sec:effect-oriented-trafos-practice}) and relate effect-oriented graph transformations to other graph transformation approaches (Sect.~\ref{sec:formal-relations-to-other-approaches}). 

\subsection{Benefiting from Effect-oriented Graph Transformation}
\label{sec:effect-oriented-trafos-practice}

There are several application cases in the literature where graph transformation has been used to achieve certain states. 
In the following, we recall graph repair, where a consistent graph is to be achieved, and model synchronisation, where consistent model relations are to be achieved after one of the models has been changed. 
A slightly different case is service matching, where a specified service should be best covered by descriptions of existing services. 

In their rule-based approach to {\em graph repair}~\cite{SH19}, Sandmann and Habel repair (sub-)conditions of the form $\exists (a\colon B \hookrightarrow C)$ using the (potentially large) set of rules that, for every graph $B^{\prime}$ between $B$ and $C$, contains the rule that creates $C$ from $B^{\prime}$. 
Negative application conditions (NACs) ensure that the rule with the largest possible $B^{\prime}$ as LHS is selected during repair. 
With effect-oriented graph transformation, the entire set of rules derived by them can be represented by a single effect-oriented rule that has the identity on $B$ as base rule and $B \hookrightarrow C$ as maximal rule. 
Moreover, we do not need to use NACs, since locally complete matching achieves the desired effect. 

In the context of model synchronisation, a very similar situation occurs in~\cite{OGLE09}. 
There, Orejas et al. define consistency between pairs of models via \emph{patterns}. 
For synchronisation, a whole set of rules is derived from a single pattern to account for the different ways in which consistency might be restored (i.e., to create the missing elements). 
Again, NACs are used to control the application of the rules. 
As above, we can represent the whole set of rules as a single effect-oriented rule. 

Fritsche, Kosiol, et al. extend {\em TGG-based model synchronisation processes} to achieve higher incrementality using special repair rules~\cite{FKST21,Fritsche22,Kosiol22}. 
Elements to be deleted according to these rules may be deleted for other reasons during the synchronisation process, destroying the  matches needed for the repair rules. 
In~\cite{FKST21}, this problem is avoided by only considering edits where this cannot happen. 
In~\cite{Fritsche22}, Fritsche approaches this problem pragmatically by omitting such deletions on-the-fly---if an element is already missing that needs be deleted to restore consistency, the consistency has already been restored locally and the deletion can simply be skipped. 
More formally, Kosiol in~\cite{Kosiol22} presents a set of subrules of a repair rule, where the maximal one matchable can always be chosen to perform the propagation. 
This whole set of rules can also be elegantly represented by a single effect-oriented rule. 

In~\cite{Arifulina17}, Arifulina addresses the heterogeneity of {\em service specifications and descriptions}. 
She develops a method for matching service specifications by finding a maximal partial match for a rule that specifies a service. 
Apart from the fact that Arifulina allows the partial match to also omit context elements, the problem can also be formulated as finding a globally maximal match (in our sense) of the service-specifying rule in the LHS of available service description rules.

\subsection{Relations to Other Graph Transformation Approaches}
\label{sec:formal-relations-to-other-approaches}
There are several approaches to graph transformation that take the variability of the transformation context into account.  
We recall each of them briefly and discuss the commonalities and differences to effect-oriented graph transformation. 

\subsubsection{Other semantics for applying single transformation rules.}

Graph transformation approaches such as the single pushout approach~\cite{Loewe93}, the sesqui-pushout approach~\cite{CHHK06}, AGREE~\cite{CDEPR15}, PBPO~\cite{CDEPR19}, and PBPO\textsuperscript{+} rewriting~\cite{OER21} are more expressive than DPO rewriting, as they allow some kind of copying or merging of elements or (implicitly specified) side effects.
Rules are defined as (extended) spans in all these approaches. 
Therefore, they also specify sets of actions that must be executed in order to apply a rule. 
AGREE, PBPO, and PBPO\textsuperscript{+} would enable one to specify what we call potential deletions; however, in these approaches their specification is far more involved. 
None of the mentioned approaches supports specifying potential creations that can be omitted depending on the currently considered application context. 

In the \emph{Double-Pullback approach}, \enquote{a rule specifies only what at least has to happen on a system\rq{}s state, but it allows to observe additional effects which may be caused by the environment}~\cite[p.~85]{HEWC01}. 
In effect-oriented  graph transformation,  the base rule also specifies what has to happen as a minimum.
But the additional effects are not completely arbitrary, as the maximal rule restricts the additional actions. 
Moreover, these additional actions are to be executed only if the desired state that they specify does not yet exist. 
This suggests that double-pullback transformations are a more general concept than effect-oriented transformations with locally complete matching. 

\subsubsection{Effect-oriented transformations via multiple transformation rules.}
\label{sec:comparison-graph-programs-formal} 
In the following, we discuss how graph transformation concepts that apply several rules in a controlled way can be used to emulate effect-oriented transformations. 

\emph{Graph programs}~\cite{HP01,CCP22} usually provide control constructs for rule applications such as sequential application, conditional and optional applications, and application loops. 
To emulate effect-oriented graph transformations, the base rule would be applied first and only once. 
For each induced rule, we would calculate the remainder rule, which is the difference to the base rule, i.e., it specifies all actions of the induced rule that are not specified in the base rule. 
To choose the right remainder rule, we would need a set of additional rules that check which actions still need to be executed in the given instance graph. 
Depending on these checks, the appropriate remainder rule is selected and applied. 

\emph{Amalgamated transformations}~\cite{BFH85,EGHLO14} are useful when graph transformation with universally quantified actions are required. 
They provide a formal basis for \emph{interaction schemes} where a \emph{kernel rule} is applied exactly once and additional \emph{multi-rules}, extending the kernel rule, are applied as often as possible at matches extending the one of the kernel rule. 
To emulate the behaviour of an effect-oriented transformation by an interaction scheme, the basic idea is to generate the set of multi-rules as all possible induced rules. 
Application conditions could be used to control that the \enquote{correct} induced rule is applied. 

A compact representation of a rule with several variants is given by \emph{variability-based (VB) rules}~\cite{SRACTP18}. 
A VB rule represents a set of rules with a common core. Elements that only occur in a subset of the rules are annotated with so-called \textit{presence conditions}. 
A VB rule could compactly represent the set of induced rules of an effect-oriented rule, albeit in a more complicated way, by explicitly defining a list of features and using them to annotate variable parts. 
An execution semantics of VB rules has been defined for single graphs \cite{SRACTP18} and sets of variants of graphs \cite{struber2018taming}.
However, for VB rules, the concept of driving the instance selection by the availability of a match with certain properties has not been developed.

	\section{Conclusion}
	\label{sec:conclusion}
	Effect-oriented graph transformation supports the modelling of systems in a more declarative way than the graph transformation approaches in the literature.  
The specification of basic actions is accompanied by the specification of desired states to be achieved.  
Dependent on the host graph, the application of a base rule is extended to the application of an induced rule that performs exactly the actions required to achieve the desired state.  
We have discussed that effect-oriented transformations are well suited to specify graph repair and model synchronisation strategies, since change actions can be accompanied by actions that restore consistency within a graph or between multiple (model) graphs. 
We have outlined how existing approaches to graph transformation can be used to emulate effect-oriented transformation but lead to accidental complexity that effect-oriented transformations can avoid. 

In the future, we are especially interested in constructing effect-oriented rules that induce consistency-sustaining and -improving transformations~\cite{Kosiol+21}. 
Examining the computational complexity of different approaches for their matching, developing efficient algorithms for the computation of their matches, elaborating conflict and dependency analysis for effect-oriented rules, and combining effect-orientation with multi-amalgamation are further topics of theoretical and practical interest. 

\subsection*{Acknowledgements}
We thank the anonymous reviewers for their constructive feedback. 
Parts of the research for this paper have been performed while J.\,K. was a Visiting Research Associate at King\rq{}s College London. 
This work has been partially supported by the Deutsche Forschungsgemeinschaft (DFG), grant TA 294/19-1.

	\bibliographystyle{splncs04}
	\bibliography{literature}
\longv{%
	\appendix
	
	\section{Additional Explanations and Results}
	\label{sec:additional-results}
	\subsection{Connectedness Conditions for Induced Rules}
\label{sec:connectedness-conditions}
As can be seen in Example~\ref{ex:induced-rules}, some of the induced rules of an effect-oriented rule can be semantically dubious such as connecting the \textsf{Account} of a \textsf{Client} to some arbitrary, already existing \textsf{Portfolio} (that might belong to another \textsf{Client}). 
Our definition of induced rules is intentionally liberal to not exclude rules that might be of interest in some application context. 
In particular, realising graph repair as suggested by Sandmann and Habel~\cite{SH19} (compare Sect.~\ref{sec:effect-oriented-trafos-practice}) via effect-oriented rules requires such a liberal definition. 
For some applications, however, it might make sense to further restrict the kind of considered induced rules. 
Our definition of \emph{connectedness conditions} gives an example of how this can be done. 

\begin{definition}[Connectedness conditions]\label{def:connectedness-conditions}
	Given an effect-oriented rule $\roo$, an induced rule of it satisfies the \emph{weak left connectedness condition} if in the factorization $(1)$ (in Fig.~\ref{fig:construction-and-matching-induced-rule}) for every node $x \in V_{\Lc^{\prime}} \setminus \iota_L^{\mathrm{i}\prime}(V_{\Lb})$, if there is an edge $e \in E_{\Lg}$ such that $\src{\Lg}(e) = \iota_L^{\mathrm{m}\prime}(x)$ and $\tar{\Lg}(e) \in \iota_L^{\mathrm{m}\prime}(V_{\Lc^{\prime}})$ or $\tar{\Lg}(e) = \iota_L^{\mathrm{m}\prime}(x)$ and $\src{\Lg}(e) \in \iota_L^{\mathrm{m}\prime}(V_{\Lc^{\prime}})$, that edge is also part of $\Lc^{\prime}$ (i.e., has a pre-image under $\iota_L^{\mathrm{m}\prime}$). 
Analogously, it satisfies the \emph{weak right connectedness condition} if for every node $x \in V_{\Kc} \setminus \iotaKg(V_{\Kb})$, if there is an edge $e \in E_{\Rg}$ such that $\src{\Rg}(e) = \ric(x)$ and $\tar{\Rg}(e) \in \ric(V_{\Kc})$ or $\tar{\Rg}(e) = \ric(x)$ and $\src{\Rg}(e) \in \ric(V_{\Kc})$, that edge is also part of $\Kc$ (i.e., has a pre-image under $\ric$).

An induced rule satisfies the \emph{left connectedness condition} (\emph{right connectedness condition}) if for every node $x \in V_{\Lc^{\prime}} \setminus \iota_L^{\mathrm{i}\prime}(V_{\Lb})$ ($x \in V_{\Kc} \setminus \iotaKg(V_{\Kb})$) every adjacent edge $e \in E_{\Lg}$ ($e \in E_{\Rg}$) is also part of $\Lc^{\prime}$ ($\Kc$).
\end{definition}

\begin{example}
	Among the induced rules from Example~\ref{ex:induced-rules} are rules that allow connecting already existing \textsf{Accounts} and \textsf{Portfolios} or \textsf{Accounts} and \textsf{Clients}, i.e., rules that enable arbitrary reuse of elements. 
	Figure~\ref{fig:example-induced-rules} shows those of the induced rules that satisfy the \textit{weak right connectedness condition}. 
	While these rules still allow some amount of reuse that might be undesired (like connecting an existing \textsf{Portfolio} to a new \textsf{Account}), the possible reuse is greatly restricted. 
	For instance, it is not possible any longer to create new connections between already existing elements. 
	
	In our example, the only induced rules that satisfy the \emph{right connectedness condition} are the rules \textsf{ensureThatClientHasAccAndPortfolio\_V1} and \textsf{ensureThatClientHasAccAndPortfolio\_V4} from Fig.~\ref{fig:example-induced-rules}. 
	This is due to the fact that in this example all nodes that are to be potentially created are connected via edges.
\end{example}

\subsection{Further Characterising Compatible Matches}
\label{sec:compatible-matches}
The next two lemmas further characterise compatible matches. 
The first result will allow us to build matches for induced rules from a pre-match for a base rule by iteratively including mappings for individual potential deletions or creations. 
\begin{lemma}[Characterising compatibility]\label{lem:compatibility-matching}
	Given an effect-oriented rule $\rooComplete$ and a pre-match $\mb$ for its base rule, an induced rule $\rc$ of $\roo$ can be matched compatibly to $\mb$ if and only if the following holds (compare Fig.~\ref{fig:construction-and-matching-induced-rule}):
	\begin{enumerate}
		\item \emph{LHS compatibility:} There exists an injective morphism $e_1\colon \Lc^{\prime} \hookrightarrow G$, called \emph{left extension match}, such that $e_1 \circ \iota_L^{\mathrm{i}\prime} = \mb$.
	
		\item \emph{RHS compatibility:} There exists an injective morphism $e_2\colon \KcOver \hookrightarrow G$, called \emph{right extension match}, such that $e_2 \circ \iotaKg = \mb \circ \leb$.
		
		\item \emph{Applicability:} The unique morphism $\mc \colon \LcOver \to G$ that exists by the universal property of the pushout and the fact that $e_2 \circ \iotaKg = \mb \circ \leb = e_1 \circ \iota_L^{\mathrm{i}\prime} \circ \leb = e_1 \circ k_1$ is injective and satisfies the dangling condition for $\rc$. 
	\end{enumerate}
	In this case, the unique morphism $\mc$ is the match for $\rc$ such that $\mb$ and $\mc$ are compatible.
\end{lemma}

\begin{proof}
	First, if $\rc$ meets the listed requirements, by the third requirement we obtain the injective morphism $\mc\colon \LcOver \hookrightarrow G$ as a pre-match for $\rc$. 
	By assumption, $\mc$ satisfies the dangling condition for $\rc$. 
	Moreover, $\mb \models \acb$ implies $\mc \models \acc = \Shift(\iotaLc,\acb)$ by~\cite[Lemma~3.11]{EGHLO14}. 
	Hence, $\mc$ is a match for $\rc$.
	Finally, by its universal property, that morphism satisfies $\mc \circ u = e_1$ which means that $\mc \circ \iotaLc = \mc \circ u \circ \iota_L^{\mathrm{i}\prime} = e_1 \circ \iota_L^{\mathrm{i}\prime} = \mb$; i.e., $\mb$ and $\mc$ are compatible. 
	
	In the other direction, given a match $\mc$ for an induced rule $\rc$ that satisfies $\mc \circ \iotaLc = \mb$, i.e., that is compatible to $\mb$, we can define $e_1 \coloneqq \mc \circ u$ and $e_2 \coloneqq \mc \circ \lec$. 
	This means that 
	\begin{align*}
		e_1\circ \iota_L^{\mathrm{i}\prime}	& = \mc \circ u \circ \iota_L^{\mathrm{i}\prime}\\
																				& = \mc \circ \iotaLc \\
																				& = \mb
	\end{align*}
	and 
	\begin{align*}
		e_2 \circ \iotaKg	& = \mc \circ \lec \circ \iotaKg\\
											& = \mc \circ u \circ k_1\\
											& = \mc \circ u \circ \iota_L^{\mathrm{i}\prime} \circ \leb\\
											& = \mb \circ \leb, 
	\end{align*}
	i.e., $\rc$ meets LHS and RHS compatibility. 
	Moreover, replacing the third equality in the computation above, this computation also shows $e_2 \circ \iotaKg = e_1 \circ k_1$. 
	This makes $\mc$ the unique morphism whose existence is implied by the universal property of the pushout computing $\LcOver$. 
	Since $\mc$ is a match by assumption, it is injective and satisfies the dangling condition for $\rc$. 
	Thus, $\rc$ satisfies all three requirements. \qed 
\end{proof}

Next, we characterise the relation between LHS and RHS compatibility and the applicability condition in the above lemma. 
In particular, this allows us to argue that only the matching of potential deletions can lead to backtracking in Algorithm~\ref{alg:matching-induced-rules}. 

\begin{lemma}[Guaranteeing applicability]\label{lem:guarantee-applicability}
	Given an effect-oriented rule $\rooComplete$, a pre-match $\mb\colon \LbOver \hookrightarrow G$ for its base rule and left and right extension matches $e_1\colon \Lc^{\prime} \hookrightarrow G$ and $e_2\colon \KcOver \hookrightarrow G$ for an induced rule that satisfy the LHS and RHS compatibility condition, respectively. 
	Then, the induced unique morphism $\mc\colon \LcOver \to G$ is injective and satisfies the dangling-edge condition for $\rc$ if and only if (i) $e_1$ and $e_2$ only identify elements that already stem from $\KbOver = \KgOver$, i.e., if $e_1(x) = e_2(y)$ implies that elements $x$ and $y$ have a common pre-image in $\Kb$, and (ii) $e_1$ satisfies the dangling-edge condition for the morphism $k_1$, i.e., if 
	\begin{equation}\label{eq:dangling-condition}\small
		\{v \in V_{L_{\mathrm{i}}^{\prime}} \mid \exists e \in E_G\setminus e_1(E_{L_{\mathrm{i}}^{\prime}}).\src{G}(e) = e_1(v) \text{ or } \tar{G}(e) = e_1(v)\} \subseteq k_1(V_{\Kb}) .
	\end{equation}
\end{lemma}

\begin{proof}
	We argue separately for injectivity and the dangling-edge condition; compare also Fig.~\ref{fig:construction-and-matching-induced-rule} for the following proof. 
	
	If $\mc$ is injective, both morphisms $e_1 = \mc \circ u$ and $e_2 = \mc \circ \lec$ are injective as compositions of injective morphisms. 
	Moreover, since the pushout computing $\LcOver$ is also a pullback (in an adhesive category) and $\mc$ is injective, the outer square $e_1 \circ k_1 = e_2 \circ \iotaKg$ is also a pullback. 
	This means that $e_1(\Lc^{\prime}) \cap e_2(\KcOver) = \KbOver$. 
	
	If, on the other hand, $e_1(\Lc^{\prime}) \cap e_2(\KcOver) = \KbOver$, this means that the outer square $e_1 \circ k_1 = e_2 \circ \iotaKg$ is a pullback.
	Then, by~\cite[Theorem~5.1]{LS05}, $\mc$ is injective. 
	
	With regard to applicability, the equation from the Lemma expresses the dangling-edge condition (see~\cite[Definition~3.9 and Fact~3.11]{EEPT06}). 
	We have to prove that
	\begin{equation}\label{eq:dangling-condition1}\scriptstyle
		M_1 \coloneqq \{v \in V_{L_{\mathrm{i}}^{\prime}} \mid \exists e \in E_G\setminus e_1(E_{L_{\mathrm{i}}^{\prime}}).\src{G}(e) = e_1(v) \text{ or } \tar{G}(e) = e_1(v)\} \subseteq k_1(V_{\Kb}) 
	\end{equation}
	if and only if
	\begin{equation}\label{eq:dangling-condition2}\scriptstyle
		M_2 \coloneqq \{v \in V_{\Lc} \mid \exists e \in E_G\setminus e_1(E_{\Lc}).\src{G}(e) = \mc(v) \text{ or } \tar{G}(e) = \mc(v)\} \subseteq \lec(V_{\Kc}) .
	\end{equation}
	Assuming Eq.~\ref{eq:dangling-condition1} to hold, let $v \in M_2$. 
	If $v \notin \lec(V_{\Kc})$, there exists $v^{\prime} \in V_{\Lc^{\prime}}$ such that $u(v^{\prime}) = v$ (because $u$ and $\lec$ are jointly surjective as co-projections of a pushout). 
	By $e_1 = \mc \circ u$ this means that $v^{\prime} \in M_1$; in particular, $v^{\prime} \in k_1(V_{\Kb})$. 
	This contradicts $v \notin \lec(V_{\Kc})$.
	Hence, $M_2 \subseteq \lec(V_{\Kc})$.
	
	In the other direction, assuming Eq.~\ref{eq:dangling-condition2} to hold, let $v \in M_1$. 
	If $u(v) \in M_2$, Eq.~\ref{eq:dangling-condition2} implies $u(v) \in \lec(V_{\Kc})$, which implies $v \in k_1(V_{\Kb})$ since the pushout computing $\Lc$ is also a pullback. 
	If $u(v) \notin M_2$, there is an edge $e^{\prime} \in E_{\Lc}$ with $\mc(e^{\prime}) = e$, where $e \in E_G$ is (one of) the edge(s) that causes $v \in M_1$. 
	Since $u$ and $\lec$ are jointly surjective, $e^{\prime}$ has a pre-image under one of them. 
	As $e^{\prime} \in u(E_{\Lc^{\prime}})$ would contradict the fact that $e$ causes $v \in M_1$, $e^{\prime} \in \lec(E_{\Kc})$. 
	This also causes $u(v) \in \lec(V_{\Kc})$ (because $\Kc$ is a graph and $\lec$ a homomorphism), which, as above, implies $v \in k_1(V_{\Kb})$. 
	Summarizing, Eq.~\ref{eq:dangling-condition2} implies $M_1 \subseteq k_1(V_{\Kb})$. \qed
\end{proof}

\subsection{Discussion of Matching Procedure}
\label{sec:representation-induced-rule}

In the following, we explain how a list of node mappings also represents the mapping of edges and the underlying induced rule and shortly remark on the need for backtracking in Algorithm~\ref{alg:matching-induced-rules}. 

During Algorithm~\ref{alg:matching-induced-rules}, the relevant data has the following form: 
The pre-match $\mb$ is given by the list \emph{currentMappings} of pairs of assigned nodes. 
This list represents the (current state of the) match $\mc$ and the LHS $\LcOver$ in the following way: 
The nodes of $\Lc$ are the nodes $n$ of $\Lg$ and $\Rg$ that are assigned to a node $x$ via a pair $(n,x)$ from \emph{currentMappings}. 
This assignment also defines $\mc$ on nodes. 
The edges of $\Lc$ are the edges of $\Lg$ and $\Rg$ that can be maximally included such that $\mc$ can be extended on them to an injective morphism to $G$. 

There are two possible sources for ambiguity in this convention: 
First, there could be parallel edges of the same type between the same two nodes in $\Lg$, $\Rg$ and/or $G$. 
In this case, all possible choices lead to isomorphic rules that delete/create the same number of edges of that type. 
Second, there could be a pair of nodes originating from $\Kb$ such that $\rg$ at the same time prescribes to delete and create an edge of a certain type between them. 
In that case, the potential creation and deletion of that type of edge compete for the same edges during matching. 
However, each choice of match and induced rule leads to an isomorphic result of the transformation. 
We could alternatively exclude such rules as they are not semantically meaningful (by prescribing to delete and re-create the same kind of edge at the same position). 
Finally, for each pair $(n,x)$ in \emph{currentMappings}, we know whether the node $n$ stems from $V_{\Lg} \setminus V_{\Kb}$ or from $V_{\Rg}$ (including $V_{\Kb}$). 
In the first case, $n$ is a node that is to be deleted, in the second case, $n$ is a node to be preserved by the induced rule. 
Thus, to obtain the corresponding induced rule, one only has to add the missing elements from $\Rg$ as creations.

It is important to note that in Algorithm~\ref{alg:matching-induced-rules} the need for backtracking is caused only by deleting nodes. 
Without nodes potentially to be deleted, by $V_{\Lc^{\prime}} = V_{\Lb}$, a left extension match satisfies Eq.~\ref{eq:dangling-condition} (and hence leads to a match for its induced rule by Lemma~\ref{lem:guarantee-applicability}) if and only if the pre-match $\mb$ it extends is already a match for the base rule. 
Therefore, in this case, we can check the dangling-edge condition once for $\mb$. 
If it is violated, we return \textsf{null}; otherwise, any selection of candidates during the algorithm results in a match and backtracking does not occur. 
In particular, the computation of all possible candidates (line~10) can be replaced by a search for a single candidate. 
This leads to an efficient computation of locally complete matches in this particular case.

	\section{Proofs of Main Results}
	\label{sec:additional-proofs}
	\begin{proof}[of Proposition~\ref{prop:base-rule-as-subrule}]
	Compare Fig.~\ref{fig:proof-rb-subrule} for the following proof.
	There, square $(2)$ is a pullback by assumption. 
	Furthermore, the top left square is a pullback since $k_1 = \iota_L \circ \leb$ (by assumption) and $\iota_L$ is injective (see, e.g., \cite[Lemma~1.10.(III)]{BS20}). 
	The lower left square is a pullback as pushout along an injective morphism (e.g., \cite[Remark~2.24]{EEPT06}). 
	By pullback composition (e.g., \cite[Fact~2.27]{EEPT06}), the two left squares together constitute a pullback. 
	In summary, rule $\rb$ embeds via pullbacks into $\rc$ as required.
	Finally, $\acc=\Shift(\iota_L,\acb)$ by definition. \qed
	
	\begin{figure}
		\centering
		\begin{tikzpicture}
			\matrix (m) [	matrix of math nodes,
										row sep=1.25em,
										column sep=1.25em,
										minimum width=1.25em,
										nodes in empty cells]
			{
				\LbOver				&								& \KbOver	&			& \\
											&								&					&			& \\
				\Lc^{\prime}	&								& \KbOver	&			& \RbOver\\
											&	\mathrm{(PO)}	&					&	(2)	&	\\
				\LcOver				& 							& \KcOver	&			& \RcOver \\};
			\node(d)[left of=m-1-1,node distance=4ex,isosceles triangle,draw=black,fill=lightgray,inner sep=2.5pt,label={[font=\scriptsize,label distance=1pt]180:$\acb$}]{};
			\node(e)[left of=m-5-1,node distance=4ex,isosceles triangle,draw=black,fill=lightgray,inner sep=2.5pt,label={[font=\scriptsize,label distance=1pt]180:$\acc=\Shift(\iota_L,\acb)$}]{};
			\path[-stealth]
				(m-1-1) edge [right hook->] node [left] {\scriptsize $\iota_L$} (m-3-1)
				(m-1-3) edge [left hook->] node [above] {\scriptsize $\leb$} (m-1-1)
								edge [-,double,double distance=2pt] node [left] {\scriptsize $\mathit{id}$} (m-3-3)
				(m-3-1) edge [right hook->] node [left] {\scriptsize $u$} (m-5-1)
				(m-3-3) edge [left hook->] node [below] {\scriptsize $k_1$} (m-3-1)
								edge [right hook->] node [below] {\scriptsize $\rib$} (m-3-5)
								edge [right hook->] node [right] {\scriptsize $\iotaKg$} (m-5-3)
				(m-3-5) edge [right hook->] node [right] {\scriptsize $\iota_R$} (m-5-5)
				(m-5-3) edge [left hook->] node [below] {\scriptsize $\lec$} (m-5-1)
								edge [right hook->] node [below] {\scriptsize $\ric$} (m-5-5);
		\end{tikzpicture}
		\caption{Proving $\rb$ to be a subrule of $\rc$}
		\label{fig:proof-rb-subrule}
	\end{figure}
\end{proof}

\begin{proof}[of Proposition~\ref{prop:number-induced-rules}]
	Every pair of sets of nodes $(A,B)$, where $A \subseteq V_{\Lg} \setminus V_{\Lb}$ and $B \subseteq V_{\Rg} \setminus V_{\Rb}$ leads to an induced rule. 
	This is by setting $\Lc^{\prime}$ (see Fig.~\ref{fig:construction-and-matching-induced-rule}) to be the graph that extends $\Lb$ with the nodes from $A$ and $\Kc$ to be the graph that extends $\Kb = \Kg$ with the nodes from $B$; note that the nodes in $B$ have to stem from $V_{\Rg} \setminus V_{\Rb}$ (and not from $V_{\Rg} \setminus V_{\Kg}$) to ensure that $(2)$ is a pullback. 
	There are $2^{\lvert V_{\Lg} \setminus V_{\Lb} \rvert + \lvert V_{\Rg} \setminus V_{\Rb} \rvert}$ such pairs of sets. 
	Up to isomorphism, these are also all possibilities to define $\Lc^{\prime}$ and $\Kc$ on nodes. 
	With regard to edges, the greatest flexibility in the choices of $\Lc^{\prime}$ and $\Kc$ exists if edges from $E_{\Lg} \setminus E_{\Lb}$ or $E_{\Rg} \setminus E_{\Rb}$ can be integrated into them independently of the choice of nodes. 
	(This is exactly the case when every such edge is incident to nodes that already stem from $\Lb$ or $\Rb$, respectively.) 
	In that case, up to isomorphism, there $2^{\lvert V_{\Lg} \setminus V_{\Lb} \rvert + \lvert E_{\Lg} \setminus E_{\Lb} \rvert + \lvert V_{\Rg} \setminus V_{\Rb} \rvert + \lvert E_{\Rg} \setminus E_{\Rb} \rvert}$ ways to construct pairs of $\Lc^{\prime}$ and $\Kc$. \qed
\end{proof}

\begin{proof}[of Proposition~\ref{prop:relations-concepts-transformation}]
	Clearly, a globally maximal transformation is also locally maximal. 
	With regard to the second statement, a rule that is not locally complete cannot be locally maximal as the existing extension leads to an applicable induced rule of larger size. \qed
\end{proof}

\begin{proof}[of Theorem~\ref{thm:characterising-eo-trafos}]
	First, let $x \in \Lg \setminus \Lb$ be an element that represents a potential deletion of $\roo$ and, if defined as a graph, let $\Kb^{+}$ be the extension of $\Kb$ with $x$ and $\iota^+\colon \Kb \hookrightarrow \Kb^+$ the corresponding inclusion. 
	Assume $m^+\colon \Kb^{+} \hookrightarrow H$ to be an injective morphism with $m^+ \circ \iota^+ = \nc \circ \iota_R \circ \rib$, where $\nc$ is the comatch of the given effect-oriented transformation $t$. 
	First, it might happen that $m^+(x)$ has a pre-image $y$ from $\Rc$ under $\nc$. 
	(This means that either $t$ first created $m^+(x)$---namely in case $y \in \Rc \setminus \Kc$---, or $y$ is a potential creation---in case $y \in \Kc \setminus \Kb$.)
	
	If $m^+(x)$ has no pre-image from $\Rc$ under $\nc$, there is an element $y \in G$ such that the so-called \emph{track morphism} of the transformation $t$ maps $y$ to $m^+(x)$ (i.e., $h(g^{-1}(y)) = m^+(x)$; compare Fig.~\ref{fig:def-rule-based-transformation}). 
	Let $\iota_L^{\mathrm{i}\prime}\colon \Lb \hookrightarrow \Lc^{\prime}$ stem from the factorisation of $\iota_L$ that leads to the computation of $\rc$ as induced rule and assume $x \notin \Lc$ (and, consequently, $x \notin \Lc^{\prime}$).
	In that situation, we can define $\Lc^{+}$ by extending $\Lc^{\prime}$ with $x$ and $e_1^{\prime}\colon \Lc^{+} \hookrightarrow G$ by extending the left extension morphism $e_1$ with the mapping $x \mapsto y$. 
	The condition $m^+ \circ \iota^+ = \nc \circ \iota_R \circ \rib$, together with $m^+(x)$ being preserved by $t$ and not having a pre-image under $\nc$, ensures that this results in an injective morphism. 
	Therefore, the factorisation of $\iota_L$ that computes $\rc$ violates the \emph{not-matchable-condition}. 
	Furthermore, Lemma~\ref{lem:guarantee-applicability} implies that it also violates the \emph{not-applicable-condition} since $y$ does not have a pre-image under the right extension match $e_2$ in this situation. 
	Summarising, assuming $x \notin \Lc$ leads to a contradiction and therefore, $x \in \Lc$. 
	(Intuitively, this means that the transformation $t$ just deleted an another possible element instead of $y$.)
	
	Secondly, let $x \in \Rc \setminus (\Kc \cup \Rb)$ represent one of the potential creations of $\roo$ that have been performed by the transformation $t$. 
	Let $\Kb^+$ be the extension of $\Kb$ with $x$ (if defined as graph), $\Kb^{\prime}$ the intersection of $\Kb^+$ and $\Kc$, and $\iota^{\prime}\colon \Kb^{\prime} \hookrightarrow \Kc$ and $\iota^+\colon \Kb^{\prime} \hookrightarrow \Kb^+$ the corresponding inclusions. 
	Assume that there exists an injective morphism $\mb^+\colon \Kb^+ \hookrightarrow G$ such that $\mb^+\circ \iota^+ = \mc \circ \lec \circ \iotaKg$. 
	Completely analogous to the procedure above, we then can extend the factorisation of $\iota_R \circ \rib$, that stems from the computation of $\rc$, by including $x$ in $\Kc$; we also extend $e_2$ by mapping $x$ to $\mb^+(x)$ in $G$. 
	In case the resulting morphism is not injective, $\mb^+(x)$ has another pre-image under $e_2$; in particular, it also has one under $\mc$. 
	In case the resulting morphism is injective, the factorisation of $\iota_R \circ \rib$ that computes $\rc$ violates the \emph{not-matchable-condition}. 
	Since $t$ is an effect-oriented transformation by assumption, this means that that factorisation satisfies the \emph{not-applicable-condition}. 
	Again, by Lemma~\ref{lem:guarantee-applicability}, this means that $\mb^+(x)$ has a pre-image under $e_1$. 
	In particular, also in this case $\mb^+(x)$ has a pre-image under $\mc$. 
	(Intuitively, this means that $\mb^+(x)$ has already been matched by some other potential action---deletion or creation.) \qed
\end{proof}

\begin{proof}[of Theorem~\ref{thm:correctness-algorithm}]
	Overall termination of Algorithm~\ref{alg:matching-induced-rules} follows from termination of the function \emph{findExtension}. 
	In this function, the for-loop iterates over a finite set of \emph{candidates} (line~12--24) and in each case the depth of backtracking is restricted by the length of the list \emph{unboundNodes} (checked in lines~14 and~26). 
	Therefore, \emph{findExtension} terminates. 
	
	With regard to correctness, first \emph{findExtension} extends \emph{currentMappings} without ever changing the beginning of that list, which stores the pre-match $\mb$ for the base rule. 
	Therefore, if Algorithm~\ref{alg:matching-induced-rules} returns a match, it is compatible with $\mb$ by construction (assuming the correctness of \emph{findExtensionCandidates} (called in line~10)). 
	Furthermore, \emph{findExtension} implements a standard recursive scheme for depth-first search. 
 	It tries every possible way to find a match that satisfies the dangling-edge condition. 
	Therefore, \emph{findExtension}, and with that \emph{findLocallyCompleteMatch}, returns \textsf{null} if and only if no induced rule of $\roo$ can be matched compatibly with $\mb$. 
	Otherwise, it computes a match.
	Finally, the computed match is locally complete as in \emph{findExtension} potential actions are always mapped if a suitable candidate exists: the matching of a potential action can only be skipped if no candidate to which it could be mapped is available (lines~11 and~25); \emph{findExtension} returns \textsf{null} if no possibility to map a given node leads to a match (line~33). 
	This means that, throughout \emph{findExtension}, if the node $n$ at position $k$ of \emph{unboundNodes} is not part of \emph{currentMappings}, either \emph{findExtension} currently has been called for a \emph{position} $\leq k$, or the assignment of the nodes at positions $1$ till $k-1$ cannot further be extended by an assignment of $n$ without violating the condition of \emph{local completeness of additional deletions} (Definition~\ref{def:oo-transformation}). \qed
\end{proof}

}
\end{document}